\documentclass[aps,prb,showpacs,twocolumn,reprint,superscriptaddress]{revtex4-1}
\usepackage{amsmath}
\usepackage{amssymb}
\usepackage{graphicx}
\usepackage{color}
\usepackage{multirow}
\usepackage{dcolumn}
\usepackage{textcomp}
\usepackage{longtable}
\usepackage{makecell}
\usepackage{booktabs}
\usepackage{array} 
\usepackage{titlesec}

\begin{document}


\title{Interface engineering of ferroelectricity in thin films of thiophosphate  $AB$P$_2$$X$$_6$ ($A$ = Cu, Ag; $B$ = In, Bi, Cr, V; and $X$ = S, Se)}

\author {Guoliang Yu}
\affiliation{Key Laboratory for Matter Microstructure and Function of Hunan Province,
Key Laboratory of Low-Dimensional Quantum Structures and Quantum Control of Ministry of Education, School of Physics and Electronics, Hunan Normal University, Changsha 410081, China}

\author {Anlian Pan}
\affiliation{Key Laboratory for Micro-Nano Physics and Technology of Hunan Province, College of Materials Science and Engineering, Hunan University, Changsha 410082, China}

\author {Mingxing Chen}
\email{mxchen@hunnu.edu.cn}
\affiliation{Key Laboratory for Matter Microstructure and Function of Hunan Province,
Key Laboratory of Low-Dimensional Quantum Structures and Quantum Control of Ministry of Education, School of Physics and Electronics, Hunan Normal University, Changsha 410081, China}

\date{\today}

\begin{abstract}
Two-dimensional ferroelectrics (FEs) are promising in the miniaturization of memory devices with ultra-high-density data storage and low power consumption. However, many 
thiophosphate monolayers, i.e., analogs of CuInP$_2$S$_6$ and referred to as ABP$_2$X$_6$, lose ferroelectricity and instead exhibit an antiferroelectric (AFE) or 
paraelectric ordering. We propose to tune the AFE ABP$_2$X$_6$ monolayers into the FE ordering through interface engineering. The 
mechanism is that there are couplings between the charge polarizations of the ABP$_2$X$_6$ monolayers and the local dipoles as well as the induced 
electronic polarizations in the substrate which have a tendency to stabilize the FE ordering. We further perform first-principles calculations for CuInP$_2$Se$_6$ 
and CuCrP$_2$S$_6$ monolayers and their van der Waals heterostructures. We find that an AFE 
CuInP$_2$Se$_6$ monolayer becomes FE as interfaced with graphene, MoS$_2$, and h-BN monolayers. In contrast, the CuCrP$_2$S$_6$ monolayer 
remains AFE since there is a large energy difference between the AFE and FE phases. Interfacing it with a MoTe$_2$ monolayer induces a metal-insulator 
transition for the heterostructure, whereas interfacing with a polar surface MgO(111) can drive it into FE. The interfacing effect can also 
be used to manipulate the FE properties of ABP$_2$X$_6$ multilayers. We further find that the AFE-to-FE transition 
is electrically switchable in these systems. In particular, it is accompanied 
by an indirect-direct band-gap transition for the CuInP$_2$Se$_6$ monolayer. Our study offers an effective 
approach to tune the FE and electronic properties of ABP$_2$X$_6$ thin films for applications in electronics and optoelectronics.

\end{abstract}

\keywords{Interface; Transition-metal dichalcogenide monolayer; valleytronics; band structure}

\maketitle
\section{INTRODUCTION}
Ferroelectric (FE) memory devices that allow nonvolatile and fast read-write processes are promising in ultra-high-density data storage with low power consumption. 
Conventional FEs, such as perovskite oxides, lose ferroelectricity as their thickness is reduced to a few nanometers due to the depolarization 
field \cite{junquera2003,Zhong1994,Fong2004,Stengel2010,Almahmoud2010}. Therefore, preserving ferroelectricity in the monolayer limit has long been an important 
issue in the field of FEs, which is desired for miniaturization of the devices. Recent studies found that group-IV monochalcogenide monolayers exhibit 
ferroelectricity with high Curie temperatures \cite{Chang2016,Fei2016,Higashitarumizu2020}. In these systems, charge polarizations 
are in-plane, which are merely affected by the depolarization field perpendicular to the systems. Notably, recent studies found that $\alpha$-In$_2$Se$_3$ monolayer 
and CuInP$_2$S$_6$ thin films possess out-of-plane polarizations \cite{Ding2017,Zhou2017,Cui2018,Liu2016,Belianinov2015}.

The above discoveries have inspired a great number of explorations of ferroelectricity in layered two-dimensional (2D) materials \cite{Wan2017,Xu2017,Xiao2018,Lin2019,Zhong2019,XuChangsong2020,Xu2020,Ma2021,Hua2021,Hua2020}. 
In particular, a number of analogs of copper indium thiophosphate CuInP$_2$S$_6$ monolayer, which are referred to as ABP$_2$X$_6$, were predicted to exhibit ferroelectricity 
and ferromagnetism simultaneously \cite{Qi2018,Lai2019}. However, like the conventional perovskite FEs, ferroelectricity in these materials is vanishing as the 
materials approaching the monolayer limit. Instead, the antiferroelectric (AFE) phase has a lower energy than that of the FE phase as revealed by density-functional theory (DFT)
calculations \cite{Qi2018,Song2017,Hua2021,Lai2019,Sun2019,Feng2020,Huang2021}. Moreover, experiment finds coexistence of FE and AFE states in CuInP$_2$Se$_6$ \cite{Dziaugys2020}. 
In fact, many ABP$_2$X$_6$ members suffer from the problem that the FE ordering is no longer the ground state (see Table~\ref{table1}); 
that is, the paraelectric (PE) or AFE ordering becomes energetically more favorable than the FE ordering. This fact is unfavorable for their applications in FE 
devices and miniaturization of the devices. Thus, it is of importance to manipulate the ferroelectricity in these systems for practical device applications.

In this paper, we find that interface can be used to tune the FE properties of ABP$_2$X$_6$ monolayers and multilayers. We provide an energy augment that involves a coupling 
between the charge polarizations of the monolayers and dipoles of the substrates. This coupling lowers the energy of one FE ordering while keeping that of the AFE ordering 
unchanged and thus may change their relative stabilities. We illustrate the idea in CuInP$_2$Se$_6$ and CuCrP$_2$S$_6$ monolayers by performing 
first-principles calculations. Moreover, we find that the CuInP$_2$Se$_6$ monolayer undergoes an indirect-to-direct band-gap transition as it transforms from the AFE ordering 
to the FE ones, which can be tuned ferroelectrically and holds potential applications in electronics and optoelectronics.

\begin{table}
\renewcommand{\arraystretch}{1.25}
\centering
\caption{Energies of ABP$_2$X$_6$ monolayers. $U_0$ ($U_1$) denotes the energy difference between the PE and FE (AFE) phases, i.e., $U_0 = E_{PE} -E_{FE}$ ($U_1 = E_{PE} - E_{AFE}$). 
$\Delta E = E_{FE} -E_{AFE}$. $E_{i}$ represent the total energies of ABP$_2$X$_6$. Data of $U_0$, $U_1$, and $\Delta E$ are not available for AgInP$_2$S$_6$ and 
AgInP$_2$Se$_6$ monolayers due to that the structures for the AFE and FE phases were relaxed to the PE phase during our DFT calculations. $GS$ denotes the ground state. 
The systems are classified into two categories by $U_0$ vs $\Delta E$.}
\label{table1} 
\begin{tabular}{lcccccc}
\toprule[0.7 pt] 
\toprule[0.7 pt]
            Systems     & $U_0$ (meV)  & $U_1$ (meV) & $\Delta E$ (meV)  &    $GS$  &   Type \\
        \hline
        CuInP$_2$S$_6$  &   218       &  242        &  24              &        AFE    &    I     \\
        CuInP$_2$Se$_6$ &   83        &   95        &  12              &        AFE    &    I     \\
        CuBiP$_2$S$_6$  &   269       &  272        &  3               &        AFE    &    I     \\
        CuBiP$_2$Se$_6$ &   121       &  104        & $-$17            &        FE     &    *     \\
        CuCrP$_2$S$_6$  &   74        &  142        &  68              &        AFE    &    I     \\
        CuCrP$_2$Se$_6$ &   20        &  69         &  49              &        AFE    &   II     \\
        CuVP$_2$S$_6$   &   67        &  139        &  72              &        AFE    &   II     \\
        CuVP$_2$Se$_6$  &  $-$7       &  71         &  78              &        AFE    &   II     \\
        AgInP$_2$S$_6$  &   N/A       &  N/A        &  N/A             &        PE     &    *     \\
        AgInP$_2$Se$_6$ &   N/A       &  N/A        &  N/A             &        PE     &    *     \\
        AgBiP$_2$S$_6$  &   44        &  3          & $-$41            &        FE     &    *     \\
        AgBiP$_2$Se$_6$ &   25        &  1          & $-$24            &        FE     &    *     \\

\toprule[0.7 pt] 
\toprule[0.7 pt]
\end{tabular}
\end{table}

\section{COMPUTATIONAL DETAILS}
We use a slab structure to model the monolayers and heterostructures. For CuInP$_2$Se$_6$-based heterostructures, the slab consists of a 1 $\times$ 1 unit cell of a CuInP$_2$Se$_6$ monolayer 
and a 2 $\times$ 2 supercell of a MoS$_2$ monolayer, whereas for CuCrP$_2$S$_6$, the slab contains a 1 $\times$ 1 unit cell of a CuCrP$_2$S$_6$ monolayer and a 
$\sqrt{3} \times \sqrt{3}$ supercell of a MoTe$_2$ monolayer. These supercells give rise to small lattice mismatches between the overlayers and the substrates ($<$ 2\%). In addition, 
for CuCrP$_2$S$_6$, a polar surface, i.e., MgO(111), is also used as the substrate. For each heterostructure, we investigate the stacking effect by performing calculations for 
a number of configurations in which the ABP$_2$X$_6$ monolayer is artificially shifted with respect to the substrate. 

We perform DFT calculations for our systems using the Vienna Ab initio Simulation Package \cite{kresse1996}. The pseudopotentials were constructed by the projector 
augmented wave method \cite{bloechl1994,kresse1999}. Van der Waals (vdW) dispersion forces between the adsorbate and the substrate were accounted for through 
the DFT-D2 method \cite{DFT-D2}. Calculations using the DFT-D3 and optPBE-vdW methods were performed for comparison \cite{klimes2010,klimes2011}. 
A 15 $\times$ 15 Monkhorst-Pack $k$-mesh was used to sample the 2D 
Brillouin zone (BZ) and a plane-wave energy cutoff of 400 eV was used for structural relaxation and electronic structure calculations. Layer projections of band structures 
were performed by using program KPROJ, which is based on the $k$-projections method \cite{Chen2014,PRB2017,PRB2018}.

\begin{figure}
  \includegraphics[width=.75\linewidth]{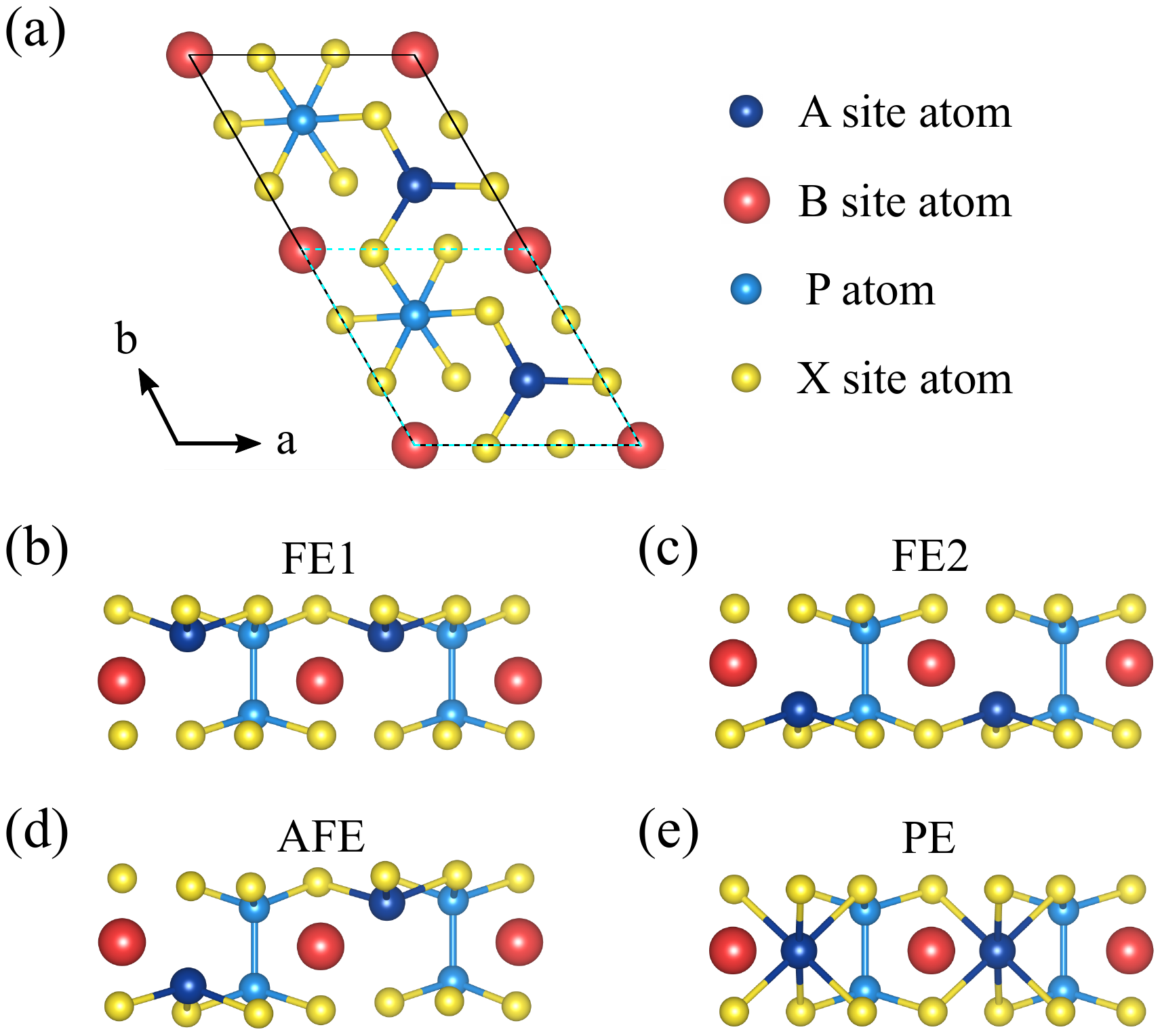}
  \caption{Geometric properties of a ABP$_2$X$_6$ monolayer. (a) Top view of the structure. The green box shows the unit cell of the FE and PE orderings. The 
  black box indicates the cell of the AFE phase, which is a $1 \times 2$ supercell of the chemical unit cell. (b), (c) Side views of the FE phases with opposite 
  charge polarizations, which are denoted as FE1 and FE2, respectively. For a free-standing ABP$_2$X$_6$ monolayer, they are energetically degenerate. (d), (e) Side views of the AFE and PE phases.}
 \label{fig1}
\end{figure}

\section{RESULTS AND DISCUSSIONS}
\subsection{General concept}
We begin by investigating the energetics of ABP$_2$X$_6$ monolayers obtained from first-principles calculations. 
We have considered PE, FE, and AFE orderings for 
all the systems, which are shown in Fig.~\ref{fig1}. In the FE and AFE phases, the charge polarizations are out of plane mainly due to displacements of $A$ atoms. 
We use a $1 \times 2$ supercell to model the AFE phase in which neighboring unit cells have opposite polarizations. The results are summarized in Table~\ref{table1}. 
One can see that for a number of ABP$_2$X$_6$ monolayers (A = Cu; B = In, Cr, V; and X = S, Se), the AFE phase has a lower energy than the PE and FE phases. For the 
FE phase, there are two energetically degenerate states, which are denoted as FE1 and FE2, respectively. In the presence of a substrate, the degeneracy is expected to be lifted. 
The trend of our results is in good agreement with previous studies \cite{Qi2018,Song2017,Hua2021,Lai2019,Sun2019,Huang2021,Feng2020}.   

\begin{figure}
  \includegraphics[width=.95\linewidth]{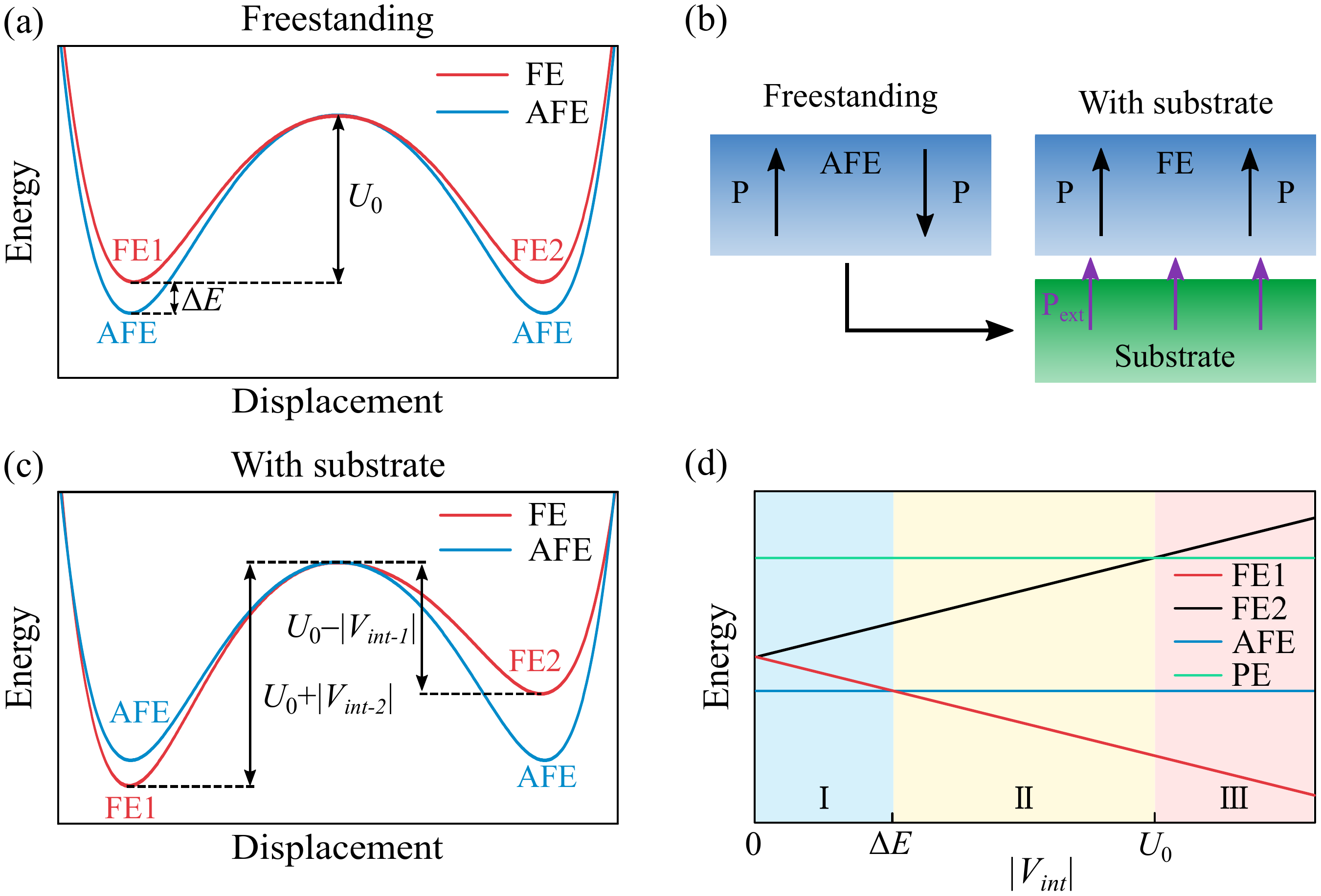}
  \caption{Interface engineering of ferroelectricity in a monolayer with out-of-plane polarization. (a) Schematic illustration of double-well potentials of a 
  free-standing monolayer for the FE and AFE phases. We assume that the AFE phase has a lower energy than the FE one. $U_0$ ($\Delta E$) denotes the energy difference 
  between the FE and PE (AFE) phases. (b) Interface induced an AFE-to-FE phase transition in a 2D monolayer with out-of-plane polarization. $P$ denotes the charge 
  polarization with arrows showing the direction. $P_{ext}$ represents dipoles of the substrate. (c) Potential energetics of the ABP$_2$X$_6$ monolayers with 
  a coupling between $P$ and $P_{ext}$. The coupling strengths $V_{int}$ for the two FE states and the substrate are expected to be slightly different. (d) 
  Total energies of different phases as a function of $V_{int}$, i.e., the coupling between $P$ and $P_{ext}$, and the phase diagram for $\Delta E < U_0$. }
 \label{fig2}
\end{figure}

We now turn to our idea of interface engineering of ferroelectricity in ABP$_2$X$_6$ monolayers, which is shown in Fig.~\ref{fig2}. The idea is pretty much 
similar to FE materials under external electric fields that the polarizations tend to be parallel to the fields. Instead, we make use of dipoles from substrates. 
The difference between using the electric fields and substrates is that the latter allows a nonvolatile tuning of the phase stability. Figure~\ref{fig2}(a) 
schematically shows the potential energy of a free-standing ABP$_2$X$_6$ monolayer as a function of displacement of $A$ atoms. 
The FE and AFE phases are expected to have double-well-like potentials. For free-standing monolayers, the two FE phases are energetically degenerate. We focus 
on the systems for which the AFE phase is the ground state. Therefore, the AFE phase is assumed to have a lower energy than the FE phases. We denote the energy 
difference between the PE and FE phases as $U_0 = E_{PE} - E_{FE}$. Generally, $U_0$ is the energy barrier for the transform between the two FE phases. 
Likewise, $\Delta$$E$ denotes the energy difference between the FE and the AFE phases, i.e., $\Delta E = E_{FE} - E_{AFE}$. The potential energy per unit cell can be 
expanded into even terms of $P$ based on the Landau-Ginzburg formula. In the presence of external dipoles $P_{ext}$, there is an additional term that accounts 
for the coupling of $P$ and $P_{ext}$. This term, essentially a dipole-dipole interaction, has a form of $-\alpha P \cdot P_{ext}/r^3$, where $\alpha$ is a 
constant and $r$ is the distance between $P$ and $P_{ext}$. We denote this term as $V_{int}$. Then, the potential energy can be written as:
\begin{equation}
 E = A P^2 + B P^4 + C P^6 + V_{int}.
\end{equation}

Therefore, the coupling lifts the degeneracy of the two FE phases: It lowers the total energy of one FE phase while it increases the total energy of the other one. The FE phase 
 with $P$ parallel to $P_{ext}$ is expected to have a lower energy than the one with $P$ antiparallel to $P_{ext}$. In contrast, the effects of this 
coupling on the total energies of the AFE and PE phases are expected to be negligible. As a result, the coupling changes the energy differences between the FE and the AFE phases and may thus 
enhance the stability of the FE phases [Figs.~\ref{fig2}(b) and 2(c)]. In ABP$_2$X$_6$-based interfaces, two types of dipoles contribute to $P_{ext}$. One is 
ion-dipole ($P_{i}$) of the substrate such as the Mo-S dipoles in MoS$_2$. The other one is the induced electronic polarization ($P_{e}$) in the substrate owing to the interface interaction.

We classify the AFE ABP$_2$X$_6$ monolayers listed in Table~\ref{table1} into two categories according to comparisons of $U_0$ and $\Delta E$: $\Delta E < U_0$ for type-I monolayers 
and $\Delta E \geq U_0$ for type-II systems, respectively. For type-I systems, the phase diagram of an AFE monolayer as a function of $V_{int}$ is shown in 
Fig.~\ref{fig2}(d). In region I, the AFE phase remains to be the ground state since the strength of $V_{int}$ is smaller than $\Delta E$. In region II, 
the coupling overcomes $\Delta E$ but remains smaller than $U_0$, i.e., $\Delta E < |V_{int}| < U_0$. In this case, FE1 has a lower energy than the 
AFE phase and FE2 remains to be a local minimum of the potential energy [see Fig.~\ref{fig2}(c)]. Therefore, the two FE phases may be switched electrically. 
In region III, i.e., $|V_{int}| \geq U_0$, FE1 remains to be the ground state, whereas FE2 is energetically higher than the PE phase. 
The coupling strengths $V_{int}$ for the two FE states are expected to be slightly different (the distance between $P$ and 
$P_{ext}$ is slightly different). Therefore, in Fig.~\ref{fig2}(c), we use $V_{int-1}$ and $V_{int-2}$ to denote the two couplings, respectively. 

For type-II systems, the FE phases remain energetically higher than the AFE phase when $|V_{int}| < \Delta E$ 
enhancing the coupling strength such that $|V_{int}| \geq \Delta E$ can tune FE1 into the ground state. However, FE2 becomes energetically higher than the PE 
phase since $\Delta E \geq U_0$ (see Appendix B). 

\subsection{CuInP$_2$Se$_6$-based heterostructures}
We now apply the above scheme to specific materials, which are CuInP$_2$Se$_6$ and CuCrP$_2$S$_6$ monolayers, respectively. Based on the values of $U_0$ and $\Delta E$ shown in 
Table.~\ref{table1}, both are type-I AFE systems. However, there is a significant difference between them. For the CuInP$_2$Se$_6$ monolayer, $\Delta E$ is small (less than 20 meV) 
and is much smaller than $U_0$, whereas for the CuCrP$_2$S$_6$ monolayer, $\Delta E$ is relatively large and is comparable to $U_0$. We choose different transition-metal dichalcogenide (TMD) 
monolayers, graphene, and h-BN monolayer as the substrates for the convenience of computation. Although a free-standing TMD monolayer is nonpolar, the local dipoles pointing from the top and bottom chalcogen atoms to 
the transition metal atoms are expected to have different effects on the total energy of the ABP$_2$X$_6$ monolayer. This is because these two types of dipoles have 
different distances to the dipole of the overlayer. The local dipoles next to the interface will dominate the coupling between the TMD and ABP$_2$X$_6$ monolayers.

\begin{figure}
  \includegraphics[width=.95\linewidth]{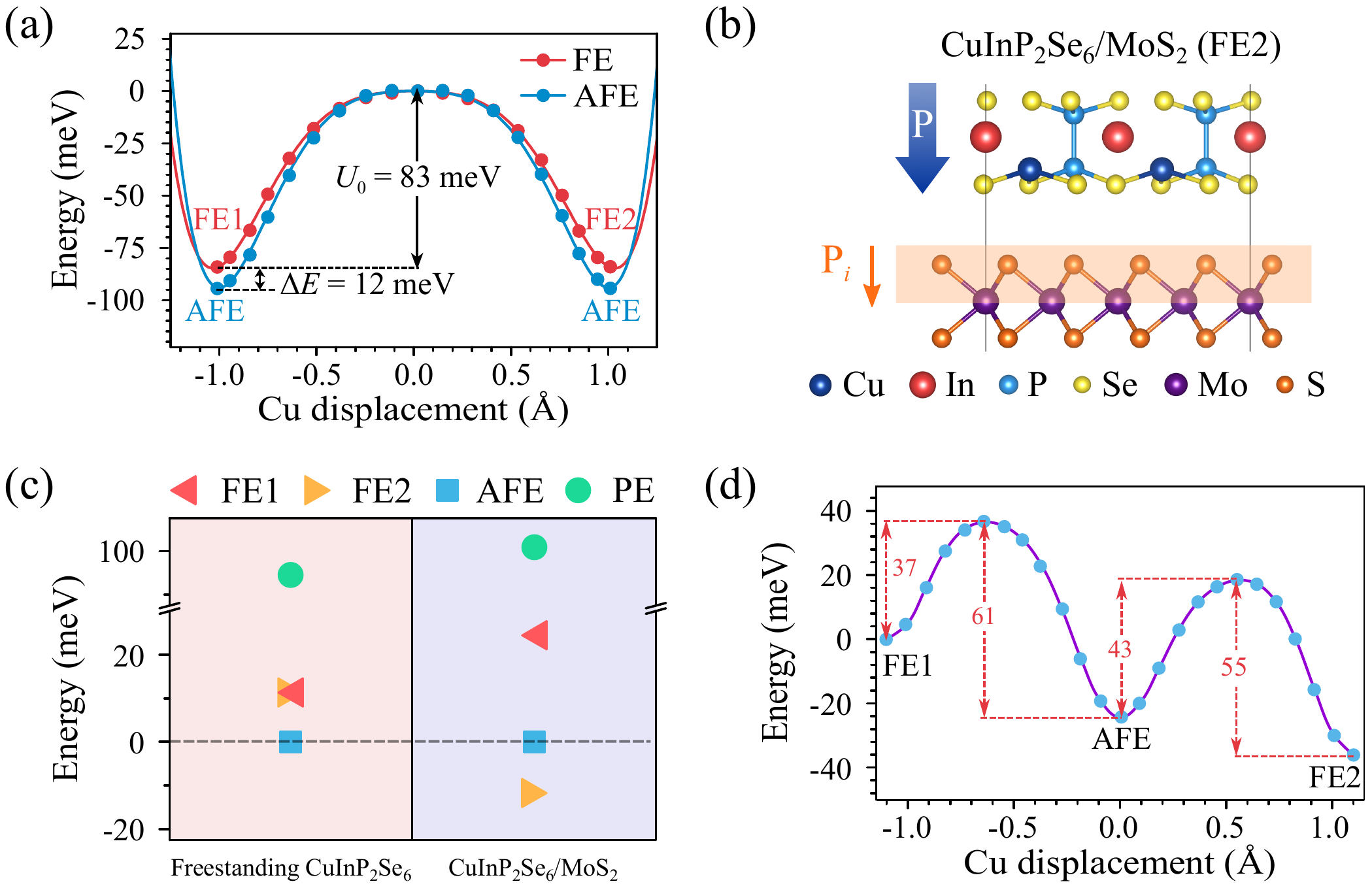}
  \caption{vdW interface engineering of ferroelectricity in CuInP$_2$Se$_6$. (a) Potential energies of the FE and AFE phases of a free-standing CuInP$_2$Se$_6$ monolayer. The 
  values of $U_0$ and $\Delta E$ are shown. (b) Structure of the FE2 phase for CuInP$_2$Se$_6$/MoS$_2$. $P$ shows the charge polarization of the CuInP$_2$Se$_6$ monolayer. 
  $P_i$ denotes the dipoles pointing from the interface S to the Mo atoms (the orange region). (c) Effects of interfacing on energies of a CuInP$_2$Se$_6$ monolayer. 
  (d) Kinetic pathway of the FE phase transform in CuInP$_2$Se$_6$/MoS$_2$. Energy barriers are shown in meV.} 
 \label{fig3}
\end{figure}

Figure~\ref{fig3}(a) shows the potential energies of a free-standing CuInP$_2$Se$_6$ monolayer.
The energy difference between the PE and the FE phases, i.e., $U_0$, is about 83 meV. This value is much larger than the energy difference between the FE and the AFE phases, 
i.e., $\Delta E$ ($\sim$ 12 meV). Such a small value of $\Delta E$ may be overcome by a vdW-type interfacial interaction between the overlayer and the substrate. We 
perform calculations of CuInP$_2$Se$_6$/MoS$_2$ for a number of configurations. For each configuration, PE, FE1, FE2, and AFE phases are considered. We find that 
the two layers interact via a vdW-type bonding since there is a large layer distance between them ($\sim$ 3.4 \AA). FE2 has the lowest energy for each configuration. 
Figure~\ref{fig3}(b) shows the side view of the lowest energy structure. The structures 
of other configurations are shown in Appendix C. From Fig.~\ref{fig3}(c), one can see that the FE2 phase is about 12 meV lower than the AFE phase. Compared to the 
free-standing system, this result indicates that a vdW interface can lead to a change of about 24 meV in the total energy of a CuInP$_2$Se$_6$ monolayer. On the other hand, the 
FE1 phase is about 25 meV higher than the AFE phase but remains about 77 meV lower than the PE phase. We further investigate the kinetic pathway of FE phase transform 
for this structure using the climbing image nudged elastic band (CI-NEB) method \cite{Henkelman2000} for which the results are shown in Fig.~\ref{fig3}(d). One can see that 
the FE2 phase remains to be a local minimum of the energy surface of CuInP$_2$Se$_6$/MoS$_2$. Moreover, the barriers between different phases suggest that a nonvolatile switching of the FE phases is feasible.

\begin{figure}
  \includegraphics[width=0.95\linewidth]{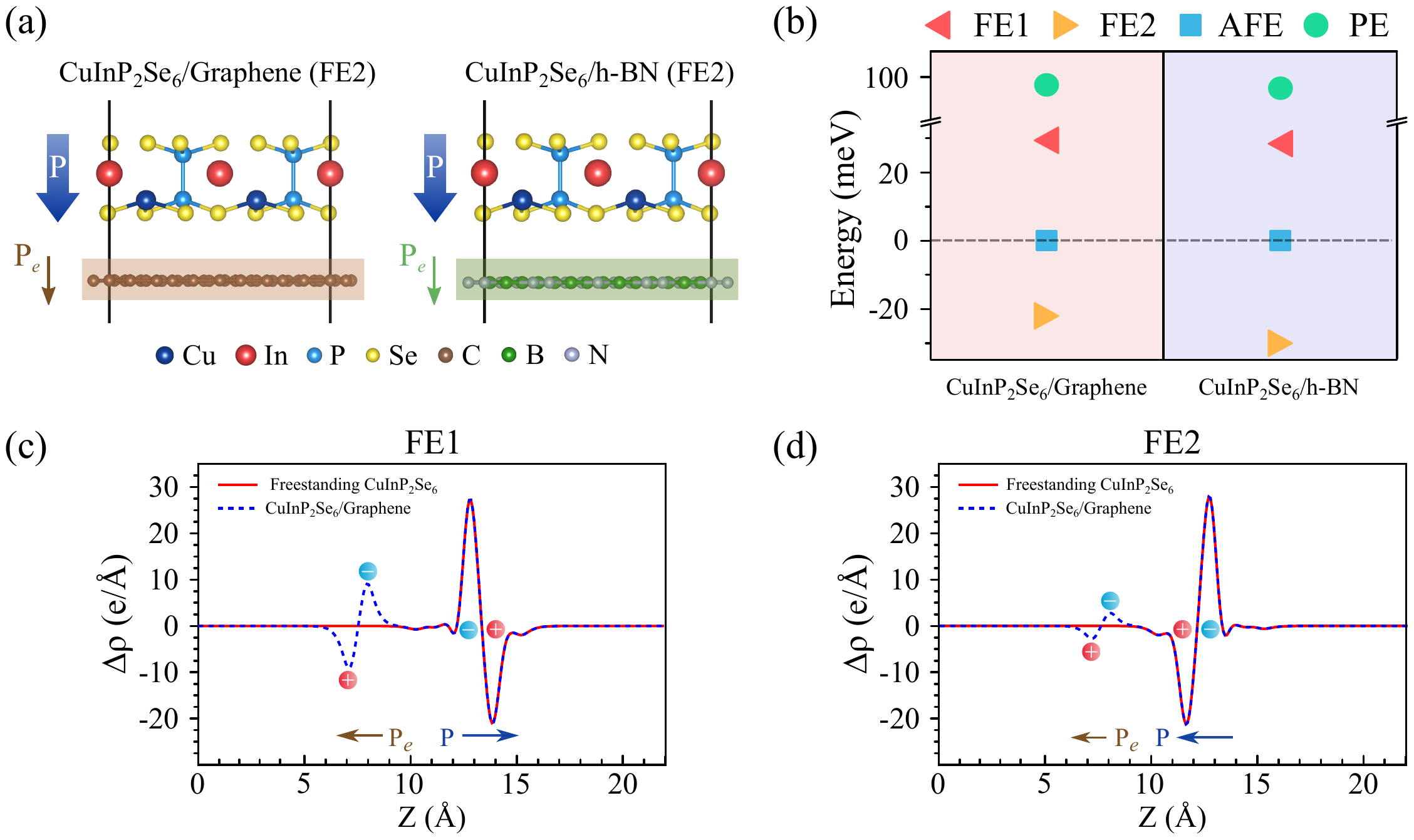}
  \caption{Ferroelectric properties of CuInP$_2$Se$_6$/graphene and CuInP$_2$Se$_6$/h-BN. (a) Geometric structure of CuInP$_2$Se$_6$/graphene and 
  CuInP$_2$Se$_6$/h-BN in FE2 phase. $P_e$ denotes the electronic polarization from interface interaction. (b) The total energies of a CuInP$_2$Se$_6$ monolayer with the substrates. (c), (d) 
  The planar-averaged charge density difference [$\Delta \rho(z)$] for the two FE orderings, which are obtained by subtracting the density of the PE phase from those of the FE orderings. 
  The red and blue spheres show the centers of the positive and negative densities, respectively. The results for CuInP$_2$Se$_6$/h-BN are similar to CuInP$_2$Se$_6$/graphene, which is shown in Appendix D.
} 
 \label{fig4}
\end{figure}

In addition to the MoS$_2$ monolayer, we also investigate the effects of interfacing the CuInP$_2$Se$_6$ monolayer to graphene as well as a h-BN monolayer [Fig.~\ref{fig4}(a)]. These 
substrates are distinct from the TMD monolayers in that they have no out-of-plane local dipoles next to the interface. Figure~\ref{fig4}(b) shows that the energy of the FE2 ordering is about 20 (30) meV
lower than the AFE phase when as the system is interfaced to graphene (a h-BN monolayer), whereas that of the FE1 phase is increased about 30 meV higher the AFE phase. 
The enhancements induced by these substrates are even slightly larger than that caused by a MoS$_2$ monolayer. We further perform analyses over the charge-density difference between the FE and PE orderings, 
i.e., $\Delta \rho = \rho_{FE} - \rho_{PE}$, from which the induced electronic polarization $P_{e}$ in the substrates can be obtained; $\rho_{FE}$ and $\rho_{PE}$ denote the charge densities of 
the FE and the PE orderings. One can see from Figs.~\ref{fig4}(c) and 4(d) 
that $P_{e}$ is antiparallel and parallel with the polarization $P$ of the FE1 and FE2 orderings of the CuInP$_2$Se$_6$ monolayer, respectively. Therefore, the coupling between $P_{e}$ and $P$ 
lowers (increases) the energy of the FE2 (FE1) phase. We also perform similar analyses for CuInP$_2$Se$_6$/MoS$_2$ and find that the electronic polarizations in the MoS$_2$ monolayer 
induced by the CuInP$_2$Se$_6$ are negligibly small (see Appendix D). These results suggest that  both the local dipoles of the substrates and the induced electronic 
polarizations in them are responsible for the enhancement of the FE ordering.

\begin{figure}
  \includegraphics[width=.65\linewidth]{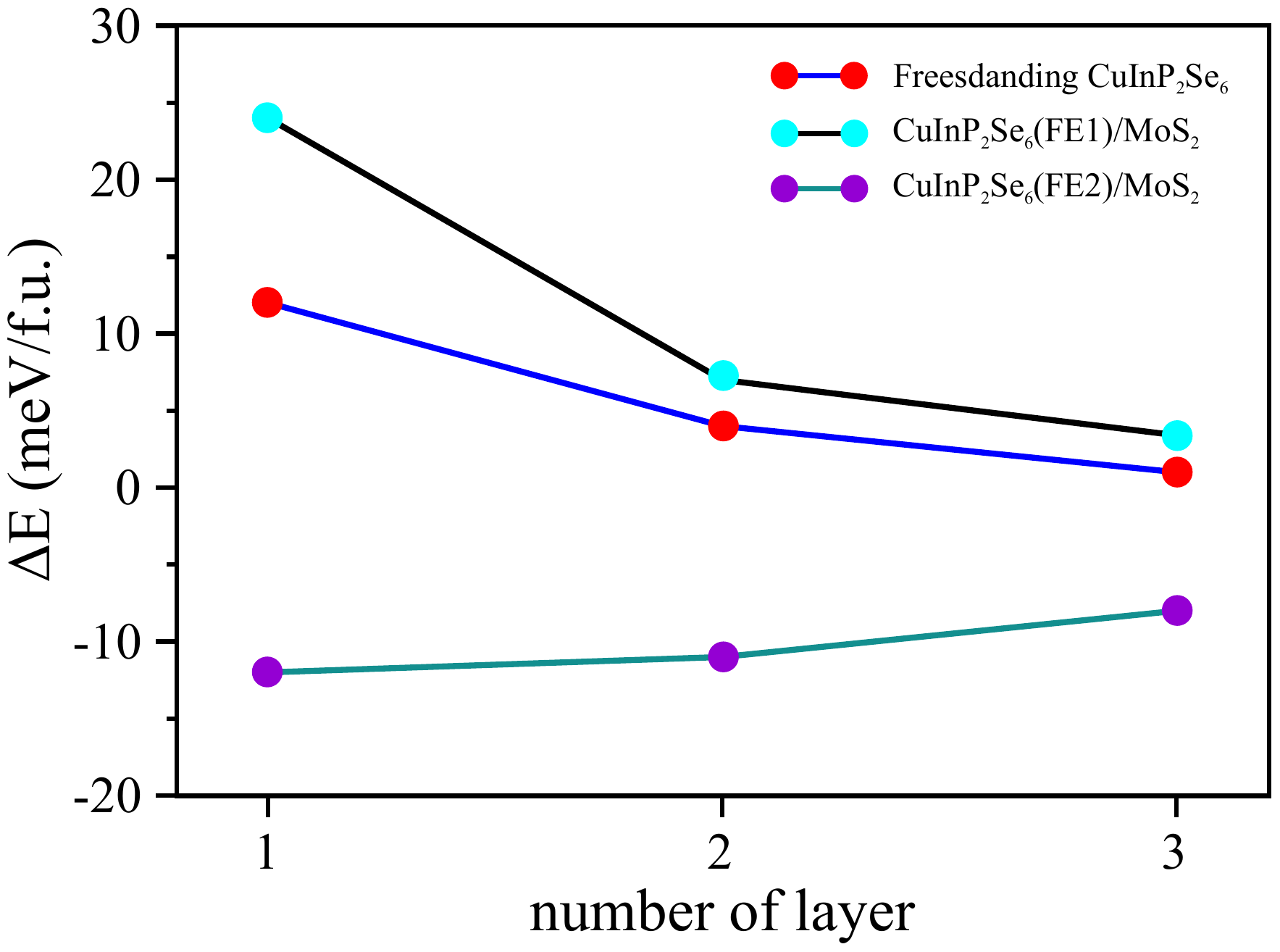}
\caption{Interfacing effects on the stabilities of various phases of CuInP$_2$Se$_6$ multilayers. $\Delta E = E_{FE} - E_{AFE}$, where $E_{FE}$ and $E_{AFE}$ are the total energies of the FE and AFE orderings. 
}
 \label{fig5}
\end{figure}

Song \textit{et al.} reveal via first-principles calculations that CuInP$_2$Se$_6$ thin films with a size of less than six layers show AFE ordering within each 
layer (the layers are ferroelectrically coupled) \cite{Song2017}. Films thicker than the critical size are found to be in the FE ordering. We perform 
calculations for its bilayer and trilayer supported by a MoS$_2$ monolayer, for which the results are shown in Fig. \ref{fig5}. One can see that both are 
turned into FE in the heterostructures. One can expect that thicker multilayers can also be driven into the FE ordering as interfaced with MoS$_2$, graphene, and h-BN monolayers.

\subsection{CuCrP$_2$S$_6$-based heterostructures}
\begin{figure}
  \includegraphics[width=.95\linewidth]{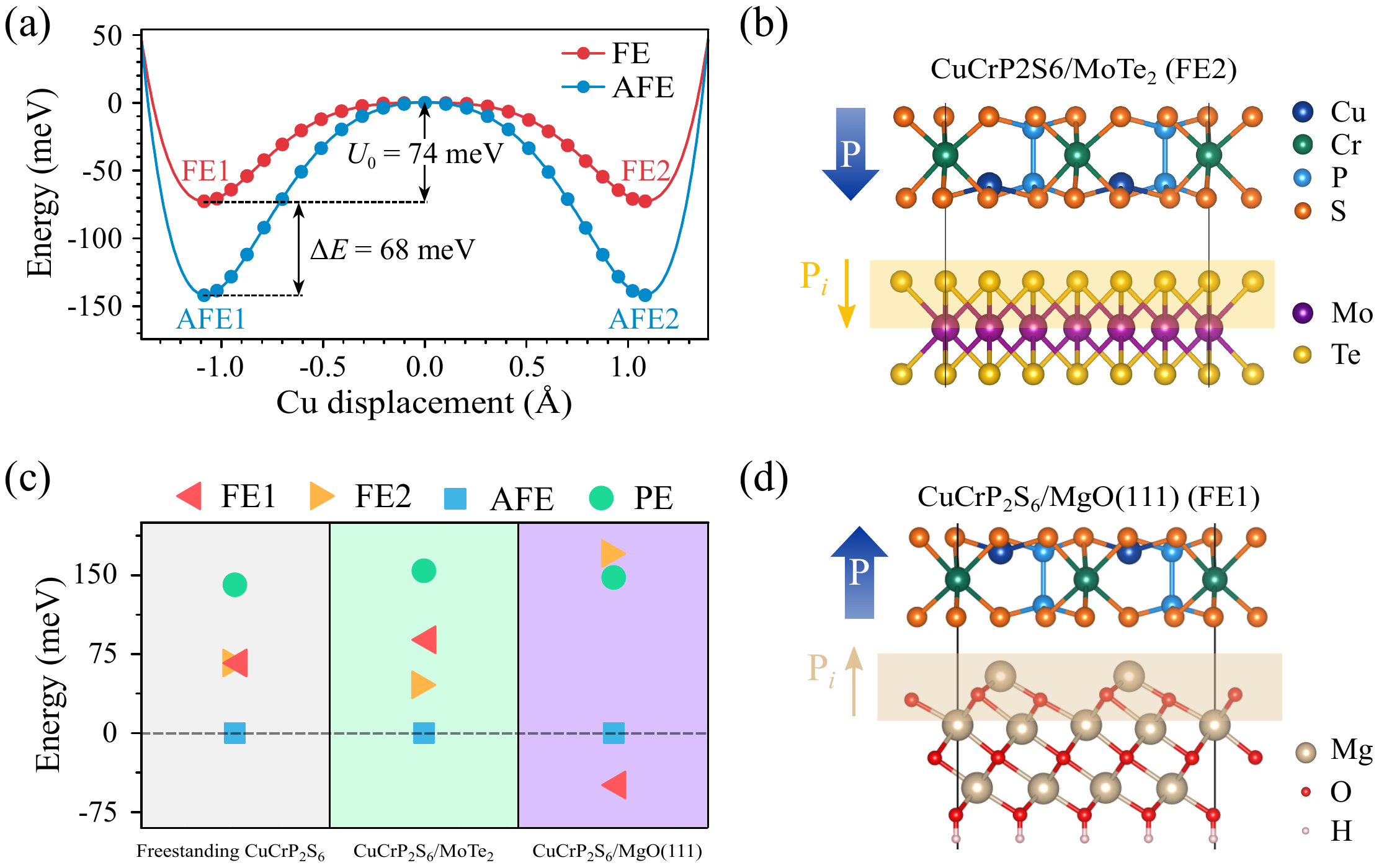}
  \caption{Interface engineering of ferroelectricity in a CuCrP$_2$S$_6$ monolayer. (a) Potential energies a free-standing CuCrP$_2$S$_6$ monolayer. (b) Geometric 
  structure of CuCrP$_2$S$_6$/MoTe$_2$ in the FE2 phase. (c) Effects of different substrates on the total energy of a CuCrP$_2$S$_6$ monolayer. (d) Geometric structure 
  of CuCrP$_2$S$_6$/MgO(111) in the FE1 ordering, respectively. In (b) and (d), charge polarizations of the ABP$_2$X$_6$ monolayer align parallel to the local dipoles 
  of the substrates next to the overlayer. } 
 \label{fig6}
\end{figure}

We now discuss the effects of interfacing on the phase stability of a CuCrP$_2$S$_6$ monolayer. The DFT+U method \cite{Dudarev1998} is used to treat 
electron correlations due to the partially filled $d$-orbital of Cr for which a value of 3 eV is used \cite{Qi2018}. From Fig.~\ref{fig6}(a) as well as Table~\ref{table1}, 
one can see that for the free-standing CuCrP$_2$S$_6$ monolayer, the energy difference between the FE and the AFE phases, i.e., $\Delta E$, is about 68 meV. This value 
is much larger than the energy change ($\sim$ 25 meV) for a CuInP$_2$Se$_6$ monolayer induced by a MoS$_2$ monolayer substrate. Geometric structures 
of the CuCrP$_2$S$_6$/MoTe$_2$ heterostructures are shown in Fig.~\ref{fig6}(b) and Appendix C. The layer distance between the two monolayers is about 3.3 \AA, which 
implies that there is also a vdW-type interaction between them. Indeed, this interaction lowers the total energy of FE2 by about 20 meV. Specifically, the energy 
difference between it and the AFE phase decreases from 68 meV for the free-standing CuCrP$_2$S$_6$ monolayer to 46 meV for the supported one. However, the AFE phase 
remains to be the ground state [see the middle panel of Fig.~\ref{fig6}(c)]. Therefore, vdW-type interfaces may not be effective in 
tuning FE properties of the CuCrP$_2$S$_6$ monolayer, although they do have a tendency to enhance the stability of the FE2 phase. 

We further look at the effects of a polar substrate on the phase stability of the CuCrP$_2$S$_6$ monolayer. We choose MgO(111) as the substrate since it is a well-known polar surface. 
Previous studies revealed that this surface demonstrates various surface reconstructions, among which octopolar and p(1 $\times$ 2) reconstructions were already experimentally 
observed \cite{Finocchi2004,Zhang2008}. In our calculations, the p(1 $\times$ 2)-MgO(111) with Mg-termination is used as the substrate for the convenience of computation. 
The structures for the heterostructures are shown in Fig.~\ref{fig6}(d) and Appendix D. Our calculations find that FE1 has a lower energy than the AFE phase by 50 meV. 
However, the total energy of FE2 is increased about 26 meV higher than the PE phase. Nonetheless, our NEB calculations shown in Appendix E suggest that there may still 
be a nonvolatile switching between the FE phases.

In addition, we perform calculations for the above CuInP$_2$Se$_6$- and CuCrP$_2$S$_6$-based interfaces using different vdW functionals/methods; the results are shown 
in Appendix F. One can see that these methods give the same trend in the stability of different orderings.

\subsection{Polarization of interfaces}
 \begin{figure}
  \includegraphics[width=0.75\linewidth]{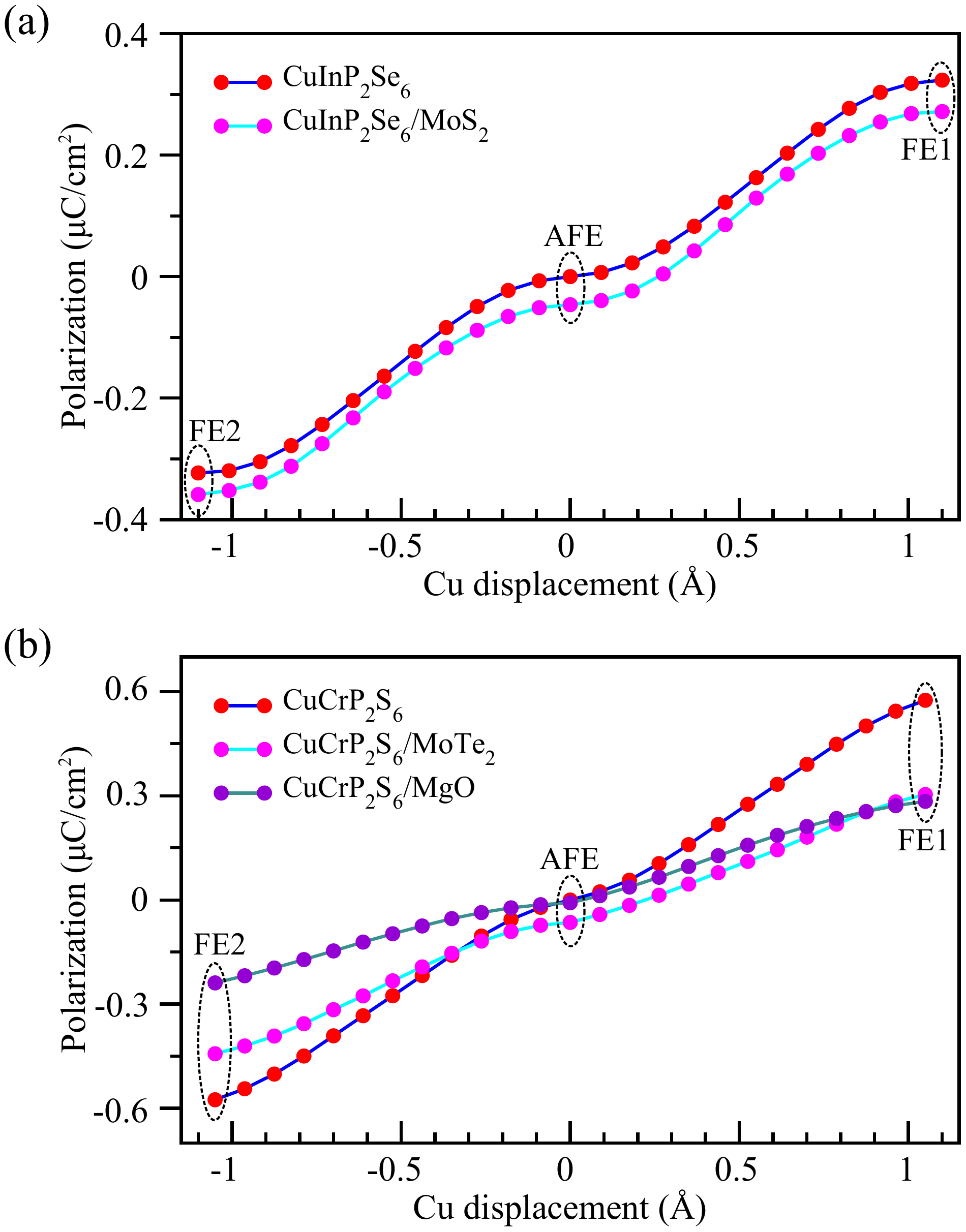}
\caption{Interfacing effects on the total polarization of CuInP$_2$Se$_6$ (a) and CuCrP$_2$S$_6$ (b). 
}
 \label{fig7}
\end{figure}

We now discuss the interfacing effects on the polarization of ABP$_2$X$_6$ monolayers. We calculate the polarization for each structure by artificially displacing 
the Cu atoms along the kinetic path FE1-AFE-FE2; the results are shown in Fig.~\ref{fig7}. The results for the free-standing CuInP$_2$Se$_6$ and CuCrP$_2$S$_6$ monolayers are 
shown for comparison. For the FE1 state of CuInP$_2$Se$_6$/MoS$_2$, the polarization is slightly smaller than that of the free-standing CuInP$_2$Se$_6$. By contrast, the polarization of 
the FE2 state is slightly enhanced as a result of interfacing. This behavior is because the induced electronic polarization $P_{e}$ is opposite to the polarization $P$ of the 
CuInP$_2$Se$_6$ monolayer for the FE1 state while parallel with $P$ for the FE2 state.

CuCrP$_2$S$_6$ shows different trend from CuInP$_2$Se$_6$ in that the polarizations for both states are reduced when it is interfaced with MoTe$_2$ and MgO(111). For the FE1 state of 
CuCrP$_2$S$_6$/MoTe$_2$, the reduction in the total polarization is because the whole system becomes metallic, which can be expected from the band alignments (see Appendix G). 
Thus, the screening effect reduces the polarization of the interface. For the FE2 state, the polarization reduction due to that the dipoles of the constituents are antiparallel with each other. For CuCrP$_2$S$_6$/MgO(111), 
the system is metallic in both the FE1 and FE2 phases due to strong interface bondings (see Appendix H). Therefore, a reduction in the total polarization can be naturally expected.

\begin{figure}[h]
  \includegraphics[width=.95\linewidth]{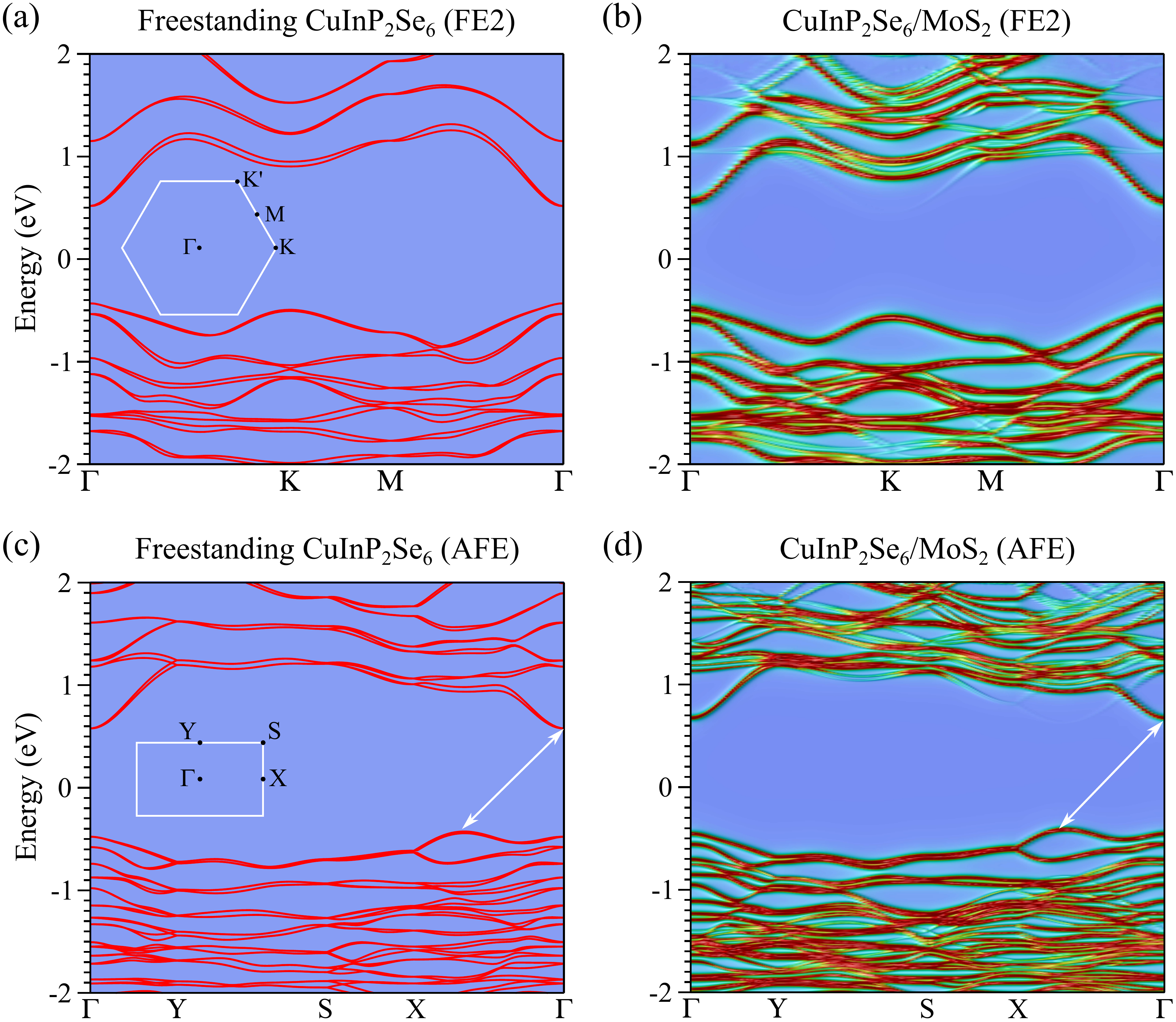}
  \caption{Effects of a MoS$_2$ monolayer substrate on the band structure of a CuInP$_2$Se$_6$ monolayer. (a) and (c) Band structures for a free-standing CuInP$_2$Se$_6$ 
  monolayer in FE and AFE orderings, respectively. (b) and (d) Corresponding plots for CuInP$_2$Se$_6$/MoS$_2$. In (b) and (d), electronic bands are weighted by 
  layer projections onto the CuInP$_2$Se$_6$ monolayer. The insets show the BZs of the CuInP$_2$Se$_6$ monolayer in different phases.} 
 \label{fig8}
\end{figure}

\subsection{Band structure}
As already mentioned above for CuCrP$_2$S$_6$/MoTe$_2$, the system experience a metal-insulator transition on a FE switch (see Fig.~\ref{fig16}). The interfacing effect can also lead to tunable 
band structure for the CuInP$_2$Se$_6$ monolayers. Figure~\ref{fig8} shows that the FE phases of a CuInP$_2$Se$_6$ monolayer 
have a direct band gap of about 0.99 eV with both the valence band maximum (VBM) and conduction band minimum (CBM) located at $\Gamma$. This band-gap size is comparable to that for Si, which is desired for optical applications.
In contrast, the AFE phase is an indirect band-gap semiconductor with a gap size of about 1.03 eV. The CBM is located at $\Gamma$, whereas the VBM is located at a $k$ point in between the $\Gamma$ and the X point.   
In the presence of a MoS$_2$ monolayer, their bands remain almost unchanged. In particular, the nature of an indirect/direct band-gap is preserved, which is due to the vdW-type 
interaction between them. We further confirm that the nature of the band gap is maintained on small strains. We recall that for 
CuInP$_2$Se$_6$/MoS$_2$ heterostructures, the FE2 phase becomes the ground state and the FE-AFE transition is ferroelectrically 
switchable. Therefore, this transition is accompanied by a switchable indirect-direct band-gap transition in the CuInP$_2$Se$_6$ monolayer, which makes the overlayer promising 
for optical and optoelectronic devices.

In conclusion, we have demonstrated that interface can be an effective way of engineering ferroelectricity in 2D ABP$_2$X$_6$ monolayers. In our model, the polarizations 
of the monolayers are coupled to local dipoles of the substrates and the induced electronic polarizations in the substrate. These couplings can stabilize the FE ordering, that is, they lower (increase) the total 
energy of one (the other) FE phase while keeping that of the AFE phase almost unchanged. We have applied the strategy to CuInP$_2$Se$_6$ and CuCrP$_2$S$_6$ monolayers, for which 
the energy difference ($\Delta E$) between the FE and the AFE phases are about 12 and 68 meV, respectively. The interfacial interaction between a CuInP$_2$Se$_6$ monolayer 
and vdW monolayers such as MoS$_2$, graphene, and h-BN leads to an energy change of 25 $\sim$ 30 meV for the FE phase and thus can overcome $\Delta E$ while keeping the other 
FE phase as a local minimum of the energy surface. Therefore, a FE switching is feasible for this system, which is supported by our NEB calculations. The vdW interfaces can 
also drive an AFE-to-FE phase transition for the multilayers of CuInP$_2$Se$_6$. However, in the case of a CuCrP$_2$S$_6$ monolayer, a vdW-type interface interaction is found to 
be insufficient to drive the FE-to-AFE phase transition. Interfacing it to a polar surface MgO(111) can make one of the FE orderings to be energetically lower than the AFE phase, 
for which a FE switching of different orderings is still likely. We further find a FE switching of electronic structure for CuInP$_2$Se$_6$ and CuCrP$_2$S$_6$-based interfaces. 
Our strategy is universal and can also be applied to other ABP$_2$X$_6$ monolayers and 2D materials with out-of-plane 
charge polarizations. Besides, our study can help understand the effects of interfacing in devices made of CuInP$_2$S$_6$ thin films and MoS$_2$ substrate as already obtained by 
recent experiments \cite{Si2018,Wang2021}.

\begin{acknowledgments}
This work was supported by the National Natural Science Foundation of China (Grants No. 11774084, No. 12174098, No. U19A2090 and No. 91833302).
\end {acknowledgments}

\section{APPENDIX}
 \begin{figure}[b]
  \includegraphics[width=.95\linewidth]{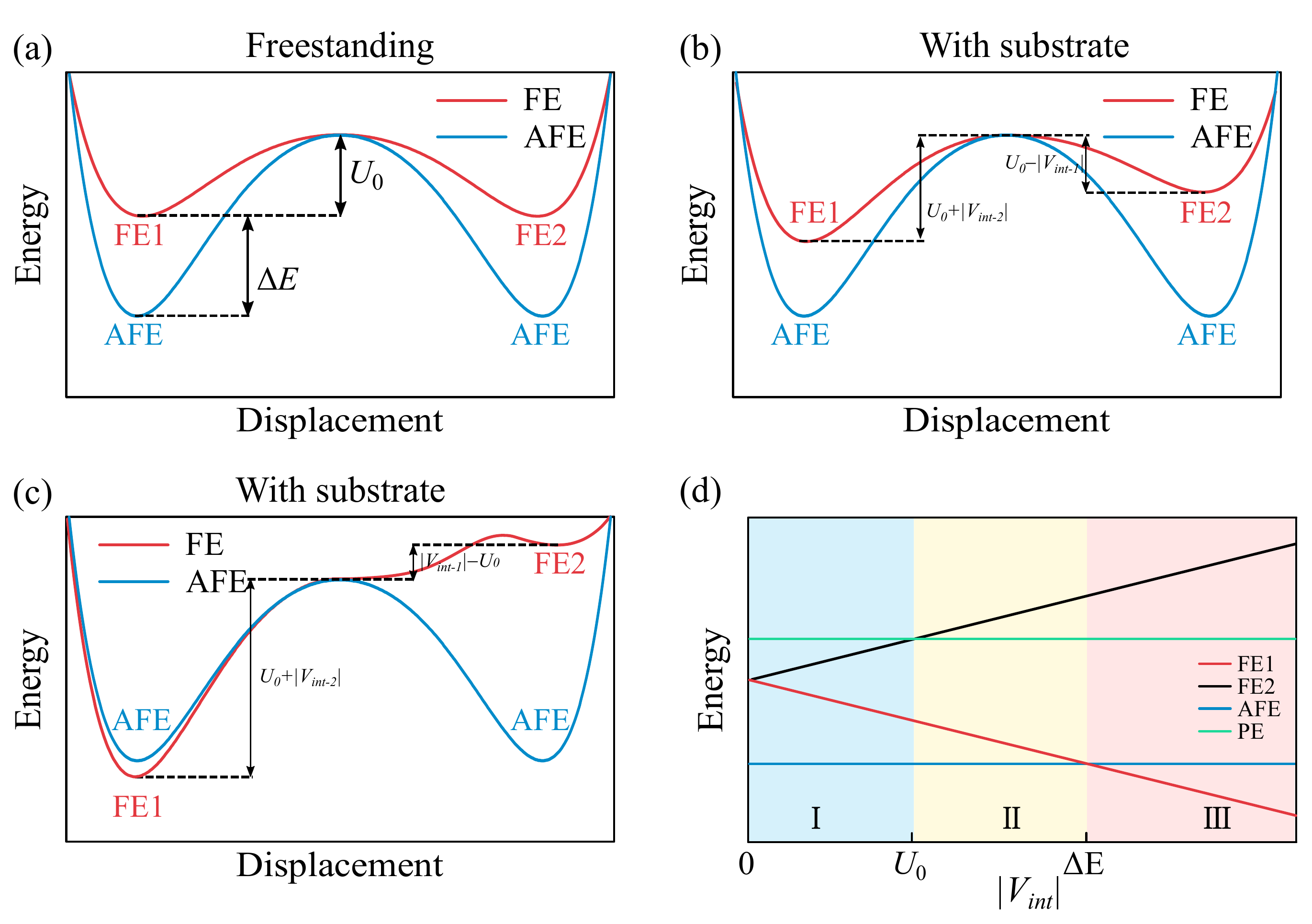}
\caption{The phase diagram of type-II systems under external dipole interactions. 
(a) Schematic illustration of double-well potentials of a free-standing monolayer for the FE and AFE phases. (b) Energy change of a ABP$_2$X$_6$ monolayer with a weak coupling 
between $P$ and $P_{ext}$ such that $|V_{int}| < \Delta E$. (c) Energy change of a ABP$_2$X$_6$ monolayer in the case of $|V_{int}| > \Delta E$. (d) The phase diagram of type-II systems as a function of $|V_{int}|$.}
 \label{fig9}
\end{figure}
 
\begin{table}
\renewcommand{\arraystretch}{1.25}
\centering
\caption{The lattice constants (\AA) of different phases for ABP$_2$X$_6$ monolayers. For AgInP$_2$S$_6$ and AgInP$_2$Se$_6$ monolayers, the structures of the AFE and FE phases were relaxed to that of the PE phase during our DFT calculations.}
\label{table2} 
\begin{tabular}{lcccc}
\toprule[0.7 pt] 
\toprule[0.7 pt]
            Systems     &     FE     & AFE        & PE      &   Ground state   \\
        \hline
        CuInP$_2$S$_6$  &   6.10     &  6.10      &  6.05   &        AFE       \\
        CuInP$_2$Se$_6$ &   6.42     &  6.42      &  6.35   &        AFE       \\
        CuBiP$_2$S$_6$  &   6.25     &  6.28      &  6.23   &        FE        \\
        CuBiP$_2$Se$_6$ &   6.55     &  6.58      &  6.55   &        FE        \\
        CuCrP$_2$S$_6$  &   6.02     &  6.02      &  6.00   &        AFE       \\
        CuCrP$_2$Se$_6$ &   6.30     &  6.28      &  6.28   &        AFE       \\
        CuVP$_2$S$_6$   &   5.98     &  5.98      &  5.93   &        AFE       \\
        CuVP$_2$Se$_6$  &   6.30     &  6.30      &  6.28   &        AFE       \\
        AgInP$_2$S$_6$  &   N/A      &  N/A       &  6.18   &        PE        \\
        AgInP$_2$Se$_6$ &   N/A      &  N/A       &  6.48   &        PE        \\
        AgBiP$_2$S$_6$  &   6.38     &  6.38      &  6.38   &        FE        \\
        AgBiP$_2$Se$_6$ &   6.65     &  6.65      &  6.63   &        FE        \\

\toprule[0.7 pt] 
\toprule[0.7 pt]
\end{tabular}
\end{table}

\begin{figure}[b]
  \includegraphics[width=.95\linewidth]{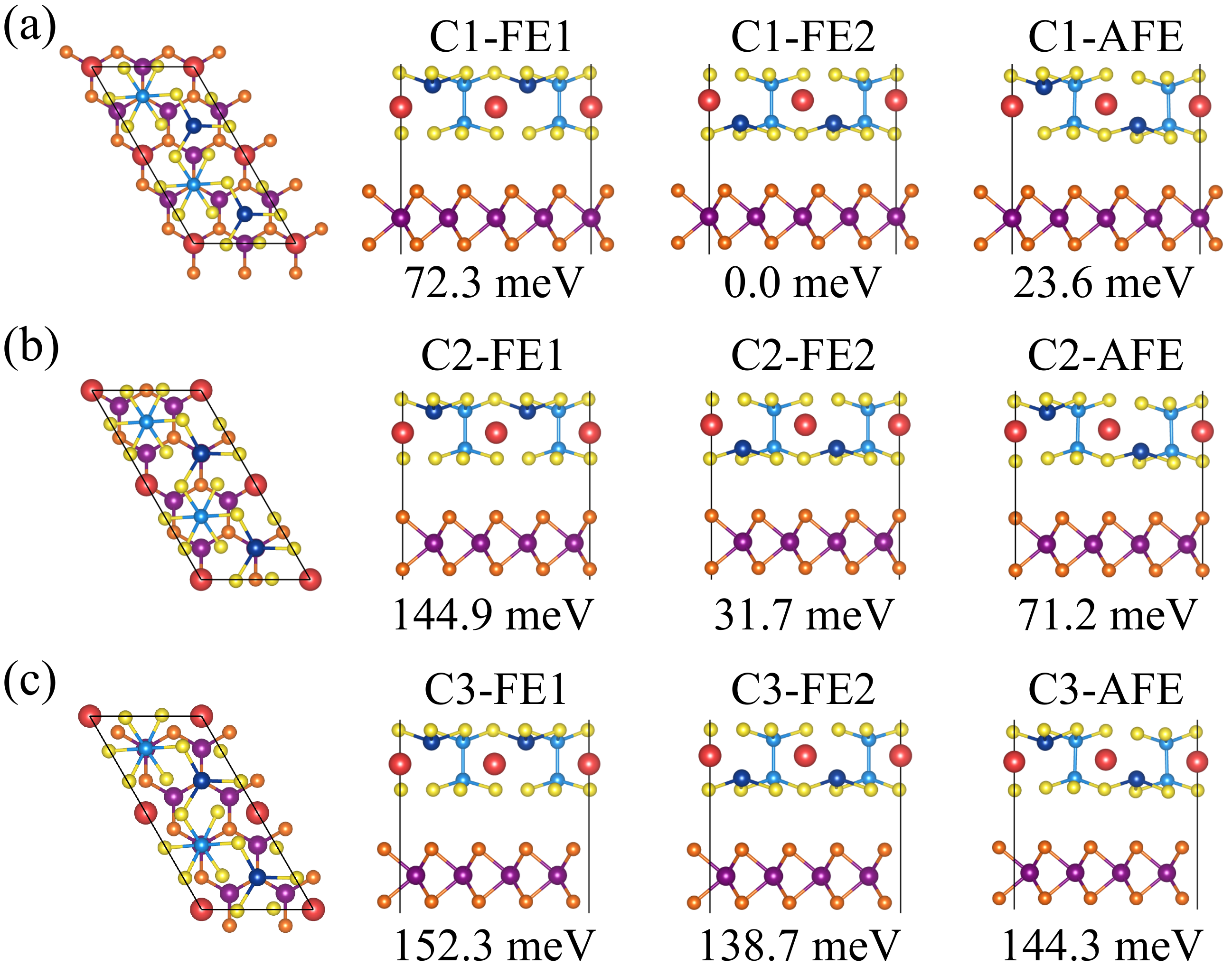}
  \caption{Top and side views of the structures of CuInP$_2$Se$_6$/MoS$_2$ with different polarization states and stackings. Here we only show three high-symmetry stackings named as  
C1, C2, and C3, respectively. The energy of each structure is given below the structure, for which that of configuration C1-FE2 is taken as the reference.} 
 \label{fig10}
\end{figure}

\subsection{Lattice constants of different phases for ABP$_2$X$_6$}
In Table~\ref{table2}, we show the optimized lattice constants of different phases for ABP$_2$X$_6$ monolayers. One can see that there are little differences in 
the lattice constant between different phases. Except for CuInP$_2$S$_6$, the ground states of the ABP$_2$X$_6$ monolayers from these calculations are consistent 
with those shown in Table~\ref{table1} obtained from calculations using the lattice constant of the FE phase for AFE and PE phases.

\subsection{The phase diagram of type-II systems}
Figure~\ref{fig9} schematically shows the phase diagram of type-II systems ($\Delta E \geq U_0$) under interactions between the ABP$_2$X$_6$ monolayers and substrates. The 
energy potentials of such a system shown in Fig.~\ref{fig9}(a) for FE and AFE phases are similar to those for type-I systems [Fig.~\ref{fig2}(a)]. The system remains AFE 
as long as $|V_{int}| < \Delta E$ [Fig.~\ref{fig9}(b)]. Strong interactions that give $|V_{int}| > \Delta E$ can turn the system into FE. However, the other FE 
phase may have an energy higher than the AFE and PE phases [Fig.~\ref{fig9}(c)]. The phase diagram shown in Fig.~\ref{fig9}(d) is different from that of type-I systems in that the 
locations of $\Delta E$ and $U_0$ are interchanged.

\subsection{Geometric structures and energetics for ABP$_2$X$_6$/Substrates}
\begin{figure}
  \includegraphics[width=.95\linewidth]{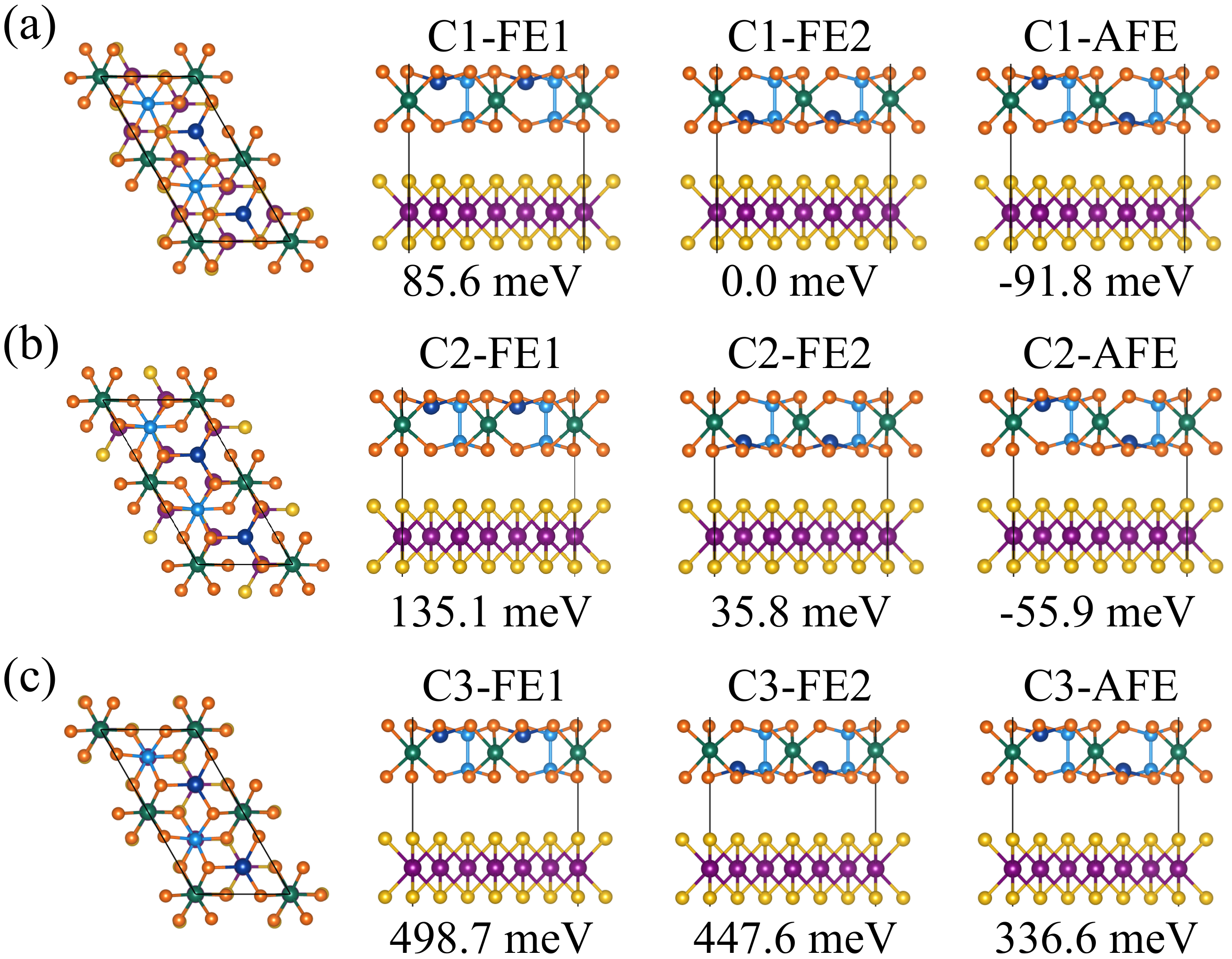}
\caption{Geometric structures of CuCrP$_2$S$_6$/MoTe$_2$ with different polarization states and stackings. The energy of each configuration relative to that of the ground state C1-FE2 is shown.}
 \label{fig11}
\end{figure}

\begin{figure}
  \includegraphics[width=.95\linewidth]{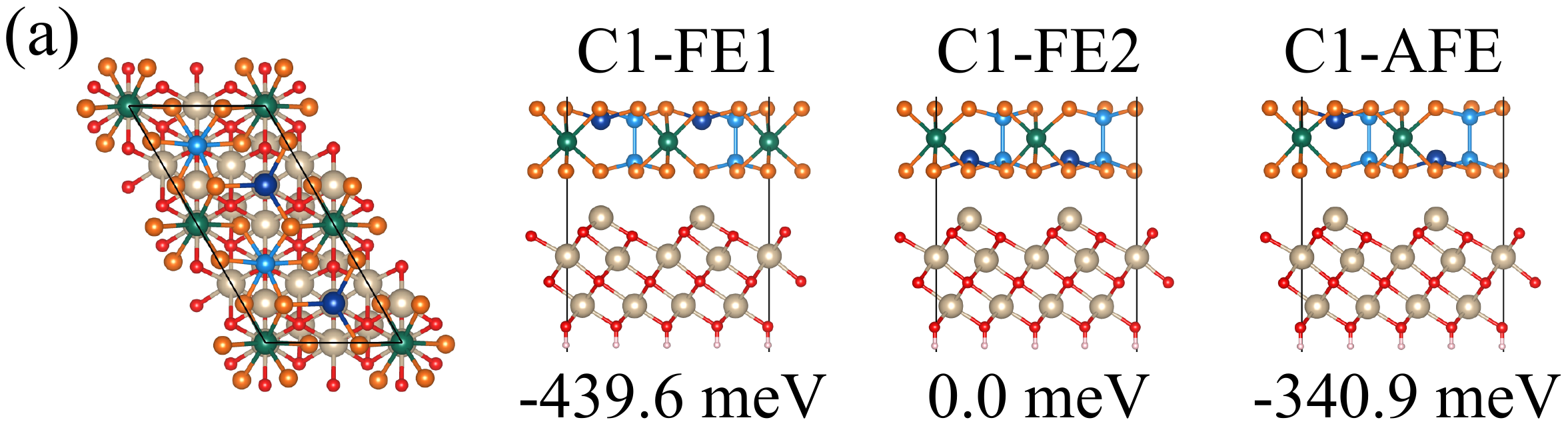}
\caption{Geometric structures of CuCrP$_2$S$_6$/MgO(111). }
 \label{fig12}
\end{figure}

Figures.~\ref{fig10}$-$\ref{fig12} show the geometric structures of CuInP$_2$Se$_6$/MoS$_2$, CuCrP$_2$S$_6$/MoTe$_2$, and CuCrP$_2$S$_6$/MgO, respectively. The energy of each structure relative 
to that of the ground state structure is also given.

\subsection{Planar-averaged differential charge density}
\begin{figure}[h]
  \includegraphics[width=.95\linewidth]{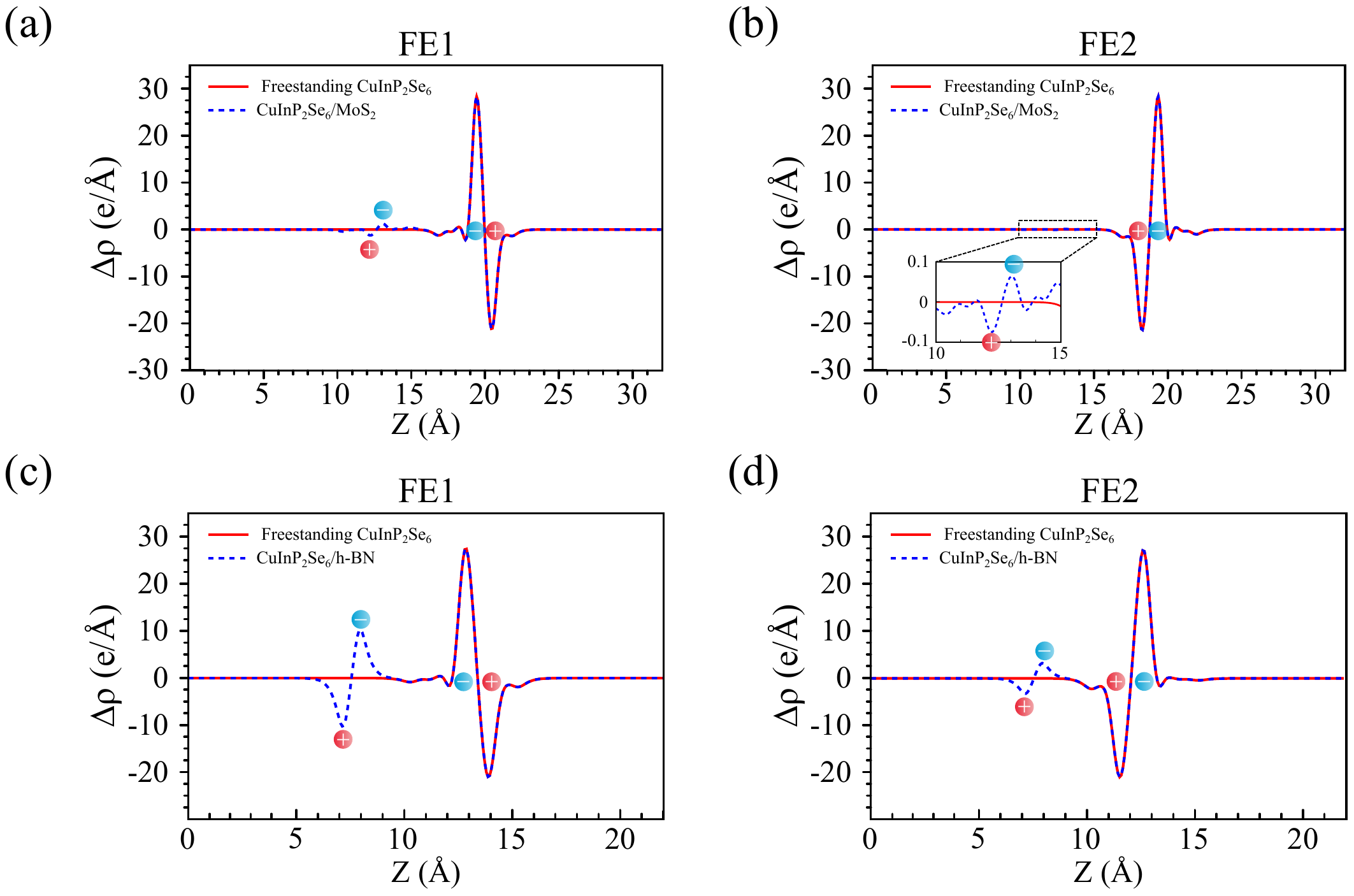}
\caption{Planar-averaged charge density difference ($\Delta \rho(z)$) for CuInP$_2$Se$_6$/MoS$_2$ and CuInP$_2$Se$_6$/h-BN in different orderings. The red 
  and blue spheres represent the centers of positive and negative charge densities, respectively.
} 
 \label{fig13}
\end{figure}

Figure~\ref{fig13} shows the planar-averaged charge density difference for CuInP$_2$Se$_6$/MoS$_2$ and CuInP$_2$Se$_6$/h-BN. One can see that the induced electronic 
polarization $P_e$ in MoS$_2$ is small for both FE orderings. By contrast, $P_e$ in h-BN is comparable to that for CuInP$_2$Se$_6$/graphene (see Fig.~\ref{fig4}).

\subsection{Kinetic pathway of the FE phase transform for CuCrP$_2$S$_6$/MgO(111)}
\begin{figure}[h]
  \includegraphics[width=.6\linewidth]{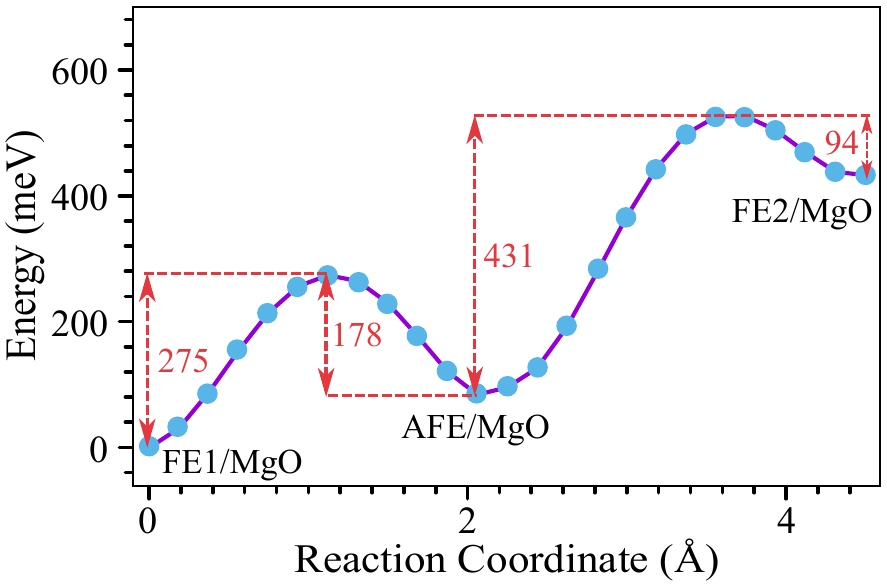}
\caption{Kinetic pathway of the FE phase transform in CuCrP$_2$S$_6$/MgO(111). Energy barriers are shown in meV.
}
 \label{fig14}
\end{figure}

Figure~\ref{fig14} shows the kinetic pathway of FE phase transform for CuCrP$_2$S$_6$/MgO(111), which indicates that a FE switching of different orderings is feasible.  

\begin{table}[b]
\renewcommand{\arraystretch}{1.25}
\centering
\caption{Energies of different orderings of ABP$_2$X$_6$ monolayers and their interfaces with different vdWs functionals/methods. The energies are calculated by taking that of the AFE phase as the reference.}
\label{table3} 
\begin{tabular}{lccccc}
\toprule[0.7 pt] 
\toprule[0.7 pt]   
            Systems                   & vdW types     & FE1      &   FE2      &    AFE     &     PE    \\
        \hline
        CuInP$_2$Se$_6$               &      *        &  12      &    12      &     0      &     95    \\
        \hline
        CuInP$_2$Se$_6$/MoS$_2$       &   DFT-D2      &  24      &   -12      &     0      &    101    \\
                                      &   DFT-D3      &  29      &   -63      &     0      &     60    \\
                                      &   optPBE      &  14      &   -23      &     0      &    115    \\
        \hline
        CuCrP$_2$S$_6$                &      *        &  68      &    68      &     0      &    142    \\
        \hline
        CuCrP$_2$S$_6$/MoTe$_2$       &   DFT-D2      &  92      &    46      &     0      &    151    \\
                                      &   DFT-D3      &  117     &    39      &     0      &    166    \\
                                      &   optPBE      &  92      &    60      &     0      &    127    \\
\toprule[0.7 pt] 
\toprule[0.7 pt]
\end{tabular}
\end{table}

\subsection{Calculate with different vdWs functionals}
The results shown in the main text were performed using DFT-D2 functional. In addition, we have carried out calculations using different vdWs functionals/methods for which the results are 
shown in Table~\ref{table3}. One can see that the results from different vdWs methods/functionals show essentially the same trend in the stability of different phases.

\subsection{Band alignments of CuInP$_2$Se$_6$ and CuCrP$_2$S$_6$ with the substrate}
\begin{figure}[h]
  \includegraphics[width=0.95\linewidth]{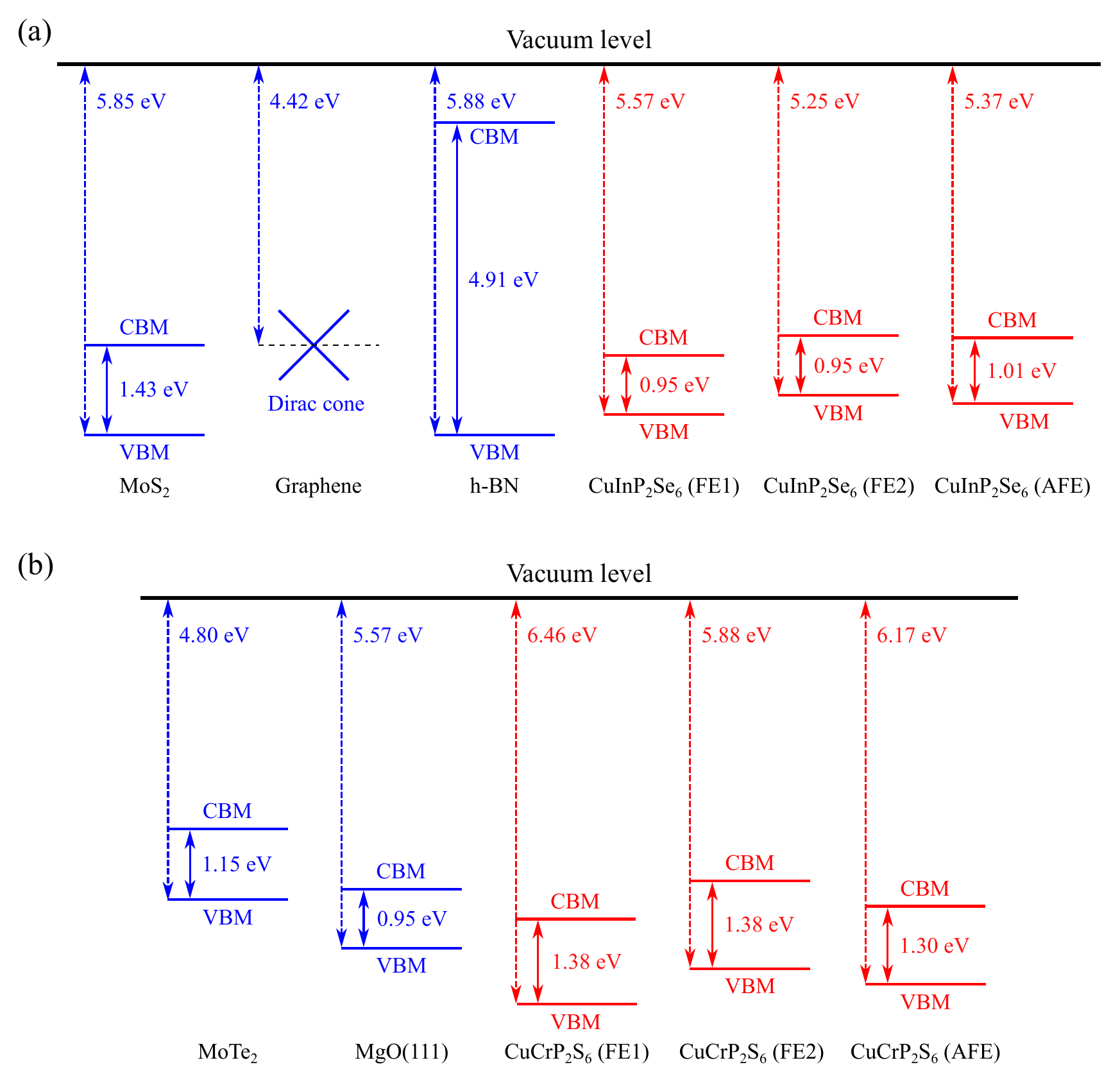}
\caption{Band alignments of CuInP$_2$Se$_6$ and CuCrP$_2$S$_6$ with the substrate. VBM and CBM denote the valence band maximum and 
the conduction band minimum, respectively. Work functions and the sizes of the bandgaps are given.
}
 \label{fig15}
\end{figure}

We investigate the workfunctions of the studied systems and their band alignments, for which the results are shown in Fig.~\ref{fig15}. One can see that CuInP$_2$Se$_6$ and MoS$_2$ as well 
as h-BN monolayers have a type-I or type-II band alignment for all the orderings. In addition, the Dirac point is located in the band gap of the FE2 and AFE phases of the CuInP$_2$Se$_6$ when their 
bands are aligned. So, charge transfers in CuInP$_2$Se$_6$/MoS$_2$, CuInP$_2$Se$_6$/graphene, and CuInP$_2$Se$_6$/h-BN heterostructures are expected to be negligibly small. 

By contrast, MoTe$_2$ has a type-III band alignment with the FE1 phase of the CuCrP$_2$S$_6$ monolayer and has a type-II band alignment with the FE2 and AFE phases. 
This behavior helps to understand the band structures shown in Fig.~\ref{fig16}. MgO(111) and the CuCrP$_2$S$_6$ monolayer has a type-II band alignment for all orderings. 
However, their heterostructures have a metallic band structure, which is due to the strong interfacial bonding.

\subsection{Band structures for CuCrP$_2$S$_6$/MoTe$_2$ and CuCrP$_2$S$_6$/MgO(111)}
\begin{figure}[h]
  \includegraphics[width=.95\linewidth]{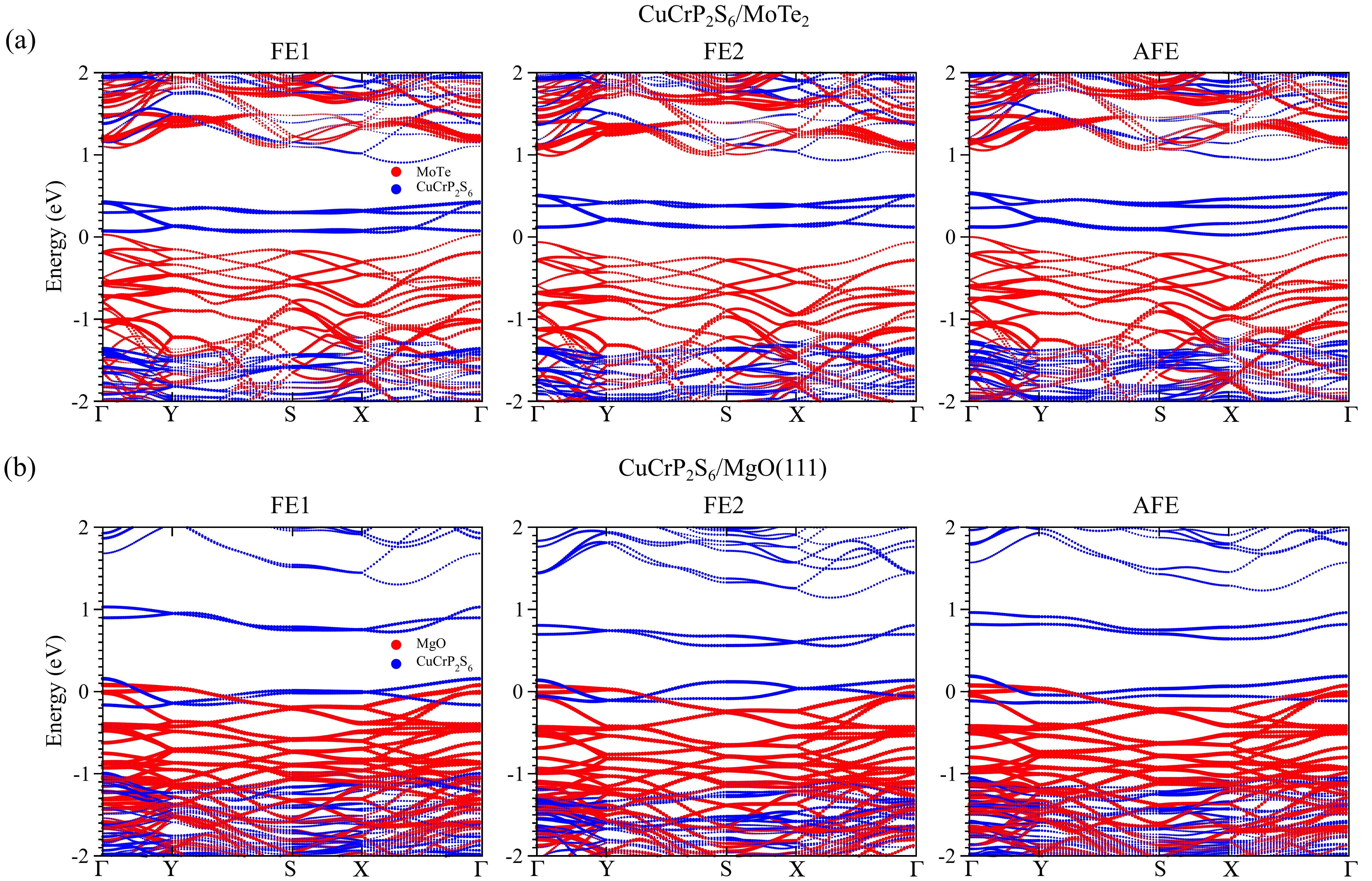}
  \caption{Effects of a MoTe$_2$ and MgO(111) substrate on the band structure of a CuCrP$_2$S$_6$ monolayer.} 
 \label{fig16}
\end{figure}

Figure~\ref{fig16} shows the band structures for CuCrP$_2$S$_6$/MoTe$_2$ and CuCrP$_2$S$_6$/MgO(111). CuCrP$_2$S$_6$/MoTe$_2$ remains semiconducting 
for the AFE and FE2 phases but becomes metallic for the FE1 phase. These behavior can be understood with the help of the band alignments 
shown in Fig.~\ref{fig15}. For CuCrP$_2$S$_6$/MgO(111), all the band structures for the three orderings indicate that the heterostructure is metallic, 
which is due to the strong interfacing bonding between the overlayer and the substrate (the interlayer distance is about 2.3 \AA).

\bibliography{references}
\bibliographystyle{apsrev4-1}

\end{document}